\documentstyle[prd,aps,eqsecnum,preprint,tighten]{revtex}
\begin{document}
\preprint{\vbox to 50 pt{\hbox{IHES/P/94/22}\hbox{CPT-94/P.E.3021}}}
\draft
%%%%%%%%%%%%%%%%%%%%%%%%%%%%%%%%%%%%%%%%%%%%%%%%%%%%%%%%%%%%%%%%%%%%%%
\title{Orbital tests of relativistic gravity\\
using artificial satellites}
\author{Thibault Damour}
\address{Institut des Hautes Etudes Scientifiques, 91440 Bures sur
Yvette, France\\
and D\'epartement d'Astrophysique Relativiste et de Cosmologie,
Observatoire de Paris,\\
Centre National de la Recherche Scientifique, 92195 Meudon, France}
\author{Gilles Esposito-Far\`ese}
\address{Centre de Physique Th\'eorique, Centre National de la
Recherche Scientifique,\\
Luminy, Case 907, 13288 Marseille Cedex 9, France}
%\date{\today}
\date{March 28, 1994}
\maketitle
%%%%%%%%%%%%%%%%%%%%%%%%%%%%%%%%%%%%%%%%%%%%%%%%%%%%%%%%%%%%%%%%%%%%%%
\begin{abstract}
We reexamine non-Einsteinian effects observable in the orbital motion
of low-orbit artificial Earth satellites. The motivations for doing so
are twofold: (i) recent theoretical studies suggest that the correct
theory of gravity might contain a scalar contribution which has been
reduced to a small value by the effect of the cosmological expansion;
(ii) presently developed space technologies should soon give access
to a new generation of satellites endowed with drag-free systems and
tracked in three dimensions at the centimeter level. Our analysis
suggests that such data could measure two independent combinations of
the Eddington parameters $\overline\beta \equiv \beta -1$ and
$\overline\gamma \equiv \gamma -1$ at the $10^{-4}$ level and probe
the time variability of Newton's ``constant" at the $\dot G/G \sim
10^{-13}$yr$^{-1}$ level. These tests would provide well-needed
complements to the results of the Lunar Laser Ranging experiment, and
of the presently planned experiments aiming at measuring
$\overline\gamma$. In view of the strong demands they make on the level
of non-gravitational perturbations, these tests might require a
dedicated mission consisting of an optimized passive drag-free satellite.
\end{abstract}

\narrowtext
%%%%%%%%%%%%%%%%%%%%%%%%%%%%%%%%%%%%%%%%%%%%%%%%%%%%%%%%%%%%%%%%%%%%%%
\section{Introduction}
\label{sec:1}

Though relativistic gravity has already been probed by quite a few
experimental tests, with the conclusion  that General Relativity is
in agreement with all existing experiments (see \cite{W92,D93} for
reviews), recent theoretical developments in tensor-scalar
cosmological models suggest that the correct theory of gravity might
differ from Einstein's theory by containing a scalar contribution
whose present magnitude has been naturally driven to a small value by
the cosmological expansion \cite{DN,DP}. These theoretical results
provide a new motivation for trying to improve the existing tests of
tensor-scalar theories of gravity.

In this paper we reexamine the use of artificial (Earth) satellites to
perform orbital tests of non-Einsteinian gravity. We shall consider here
only the most conservative deviations from Einstein theory, i.e.
boost-invariant effects associated to the exchange of scalar excitations.
[We refer to our recent work \cite{DEF94} for a discussion of the use of
satellite data as probes of possible preferred-frame effects in
relativistic gravity.] Although several of the Einsteinian or
non-Einsteinian effects in the motion of artificial satellites have been
abundantly discussed in the literature \cite{W92,S89,C89,B91}, it seems
that there exists no treatment of the problem which is both systematic
and observation-oriented. There exists a systematic analysis of
non-Einsteinian effects in the motion of the Moon \cite{N73}, but we
shall see that the two problems are quite dissimilar both because of
differences in physical parameters ($a_{\rm Moon} /R_{\rm Earth} \sim
60$ compared to $a_{\rm satellite} /R_{\rm Earth} \sim$ a few) and of
differences in observational techniques (radial ranging for the Moon
versus three-dimensional tracking for satellites). The aim of the
present paper is to provide a systematic analysis of non-Einsteinian
effects possibly observable in the orbital data of drag-free satellites
whose three-dimensional motion is tracked (either by Laser or by Global
Positioning System (GPS) techniques) at the centimeter level. The
Gravity Probe B (GPB) satellite should soon provide the first example of
such satellites. [The existing Laser Geodynamical Satellites (LAGEOS) are
tracked with centimeter precision, but the small level of residual drag
acting on their motion is probably too large for them to be useful
probes of the effects we discuss below.]

 Note that, contrary to some proposals, e.g. \cite{C89}, we shall not
consider explicitly here the purely Einsteinian relativistic effects.
We consider that the ensemble of existing positive tests of General
Relativity \cite{W92,D93} has already probed most of the qualitatively
crucial features of Einstein's theory. We shall therefore assume that
the Einsteinian post-Newtonian contributions to the satellite equations
of motion are separately included in their entirety (see \cite{DSX}),
and focus on the quantitatively most promising effects for measuring
possible non-Einsteinian weak-field deviations.

In the post-Newtonian limit, generic tensor-scalar theories of gravity
predict that the motion of $N$ spherical, non rotating bodies is given
by the Lagrangian \cite{DEF92}
\begin{eqnarray}
 L_{\overline\beta ,\overline\gamma ,\overline\delta} =&& - \sum_A m_A
c^2 (1 - {\bf v}^2_A /c^2)^{1/2} \nonumber \\
  && + {1\over 2} \sum_{A\not= B} {G_{AB}\, m_A\, m_B \over r_{AB}}
\left[ 1 + {3\over 2c^2} ({\bf v}^2_A + {\bf v}^2_B) \right. \nonumber \\
  && - {7\over 2c^2} ({\bf v}_A\cdot {\bf v}_B) - {1\over 2c^2}({\bf
n}_{AB} \cdot {\bf v}_A) ({\bf n}_{AB} \cdot {\bf v}_B) \nonumber \\
  && \left. \qquad \qquad + {\overline\gamma \over c^2} ({\bf v}_A
       - {\bf v}_B)^2  \right] \nonumber\\
  && - {1\over 2} \sum_{B\not= A\not= C} (1+2 \overline\beta) {G^2
     m_Am_Bm_C \over c^2 r_{AB} r_{AC}} \ , \label{eq:1.1}
\end{eqnarray}
where $r_{AB} \equiv |{\bf x}_A - {\bf x}_B|$, ${\bf n}_{AB} \equiv
({\bf x}_A - {\bf x}_B)/ r_{AB}$, ${\bf v}_A \equiv d{\bf x}_A /dt$,
and where the summation symbols denote multiple sums over the various
independent body labels $A,B$,\dots with the only exclusion of the would
be infinite contributions.

In Eq.~(\ref{eq:1.1}) the effective coupling constant $G_{AB}$ for the
gravitational interaction of bodies $A$ and $B$ is of the
form\footnote{In Eq.~(\ref{eq:1.2}), the basic gravitational constant $G$
is the sum of the contribution due to the tensor interaction, and of a
mean contribution due to the scalar interaction. One assumes that the
small composition-dependent effects $\overline\delta_A \ll 1$, and
neglects terms of order $\overline\delta^2$.}
\begin{equation}
 G_{AB} = G [ 1 + \overline\delta_A + \overline\delta_B ] \label{eq:1.2}
\end{equation}
where $\overline\delta_A$ is generically the sum of two physically
independent contributions
\begin{equation}
 \overline\delta_A = \widehat\delta_A + \eta\ {E^{\rm
grav}_A\over m_Ac^2}\ ,
 \label{eq:1.3}
\end{equation}
with
\begin{equation}
 \eta \equiv 4\, \overline\beta - \overline\gamma \ . \label{eq:1.4}
\end{equation}

The first contribution $\widehat\delta_A$ comes from the
fact that the physically best motivated tensor-scalar theories violate
the weak equivalence principle \cite{DP}, while the contribution
proportional to the gravitational self-energy
 \[ E^{\rm grav}_A = - (G/2) \int_A \int_A d^3x\, d^3x'\, \rho ({\bf x})
 \rho ({\bf x}') / |{\bf x} - {\bf x'}|  \]
derives from the fact that all the theories containing a variable local
gravitational ``constant'' violate the strong equivalence principle
\cite{N68}.

The dimensionless numbers $\overline\beta$, $\overline\gamma$ parametrize
weak-field {\it deviations} from General Relativity and are proportional
to the strength (relative to the usual tensor contribution) of the
coupling to matter of the scalar contribution to gravity \cite{DEF92}.
They are related to the usual post-Newtonian (Eddington) parameters
simply by $\overline\beta \equiv \beta -1$, $\overline\gamma \equiv
\gamma -1$ (so that $\eta = 4\beta -\gamma -3$). Besides the effects
parametrized by $\overline\beta$, $\overline\gamma$ and the
$\overline\delta$'s, another deviation from General Relativity generically
predicted by tensor-scalar theories, and implicitly contained in the
Lagrangian (\ref{eq:1.1}), is a possible time variability of the
gravitational coupling strength $G$, as well as (in generic tensor-scalar
theories \cite{DP}) a time variability of the masses\footnote{The
other parameters $\overline\beta$, $\overline\gamma$,
$\overline\delta_A$ are generically also slowly changing. However, this
leads to higher-order, unobservably small effects.} $m_A$.

The best present (one sigma) observational limits on the parameters of
the Lagrangian (\ref{eq:1.1}) measuring deviations from Einstein's
theory are:
$|\overline\gamma| < 2\times 10^{-3}$ \cite{Viking},
$|\overline\beta| < 3\times 10^{-3}$ (assuming a Sun's $J_2 \sim 2 \times
10^{-7}$) \cite{S90}, $\overline\delta_{\rm Moon} -
\overline\delta_{\rm Earth} = (-2.7 \pm 6.2) \times 10^{-13}$ \cite{LLR},
and $|\dot G/G| \lesssim  10^{-11}$yr$^{-1}$ \cite{S90} (we are using the
fact that combined geophysical and astronomical data set a bound of order
$10^{-13}$yr$^{-1}$ on the possible time variability of the masses
\cite{SV90}). Note that if one were to assume the exact validity of the
weak equivalence principle, i.e. $\widehat\delta_A \equiv 0$,
the Lunar Laser Ranging result on $\overline\delta_{\rm Moon} -
\overline\delta_{\rm Earth}$ would imply $\eta =(-0.6 \pm 1.4) \times
10^{-3}$, from which ---~using the Viking limit~--- one would deduce
$\overline\beta =(-1\pm 5)\times 10^{-4}$ \cite{LLR}. However, in the
present work we have in mind the general class of tensor-scalar theories
which do not respect the weak equivalence principle (such as in
Ref.~\cite{DP}). As the present observational limits on the
$\widehat\delta_A$'s are not better than $3\times 10^{-12}$
\cite{Su}, we must consider the
$\overline\delta_A$'s in Eq.~(\ref{eq:1.1}) as being independent from
$\overline\beta$ and
$\overline\gamma$.

 The object of the present paper is to point out that the
centimeter-level tracking of drag-free artificial Earth satellites
is a promising tool for measuring {\it both} $\overline\beta$ and
$\overline\gamma$ at the $10^{-4}$ level, and $\dot G/G$ at the
$10^{-13}$yr$^{-1}$ level. We shall show that there are two independent
relativistic effects which build up over many orbital periods and lead
to displacements $\sim O (\overline\beta , \overline\gamma) \times
10^4$cm for integration times of the order of 6 months. One of these two
relativistic effects is non null and is the well known perigee advance
associated with the (Eddington-modified) Schwarzschild field of the
Earth which has been abundantly discussed in the literature
\cite{W92,S89,B91,Tapley}.  The other one is a null relativistic effect
whose origin is known \cite{W92,N73}, but whose importance, in the
context of low orbit satellites, for testing the combination $\eta =
4\overline\beta - \overline\gamma$ has (as far as we are aware) not been
clearly realized before. [See further discussion below.]
%%%%%%%%%%%%%%%%%%%%%%%%%%%%%%%%%%%%%%%%%%%%%%%%%%%%%%%%%%%%%%%%%%%%%%
\section{Lagrangian and equations for motion for satellite motion.}
\label{sec:2}

 The Lagrangian (\ref{eq:1.1}) describes the global dynamics of the
bodies of the solar system (``barycentric'' reference frame). We wish
to pass to a ``geocentric'' reference frame. A rigorous and elegant
way of effecting such a transformation in the framework of General
Relativity has been recently discussed in Ref.~\cite{DSX}. However,
we are here working within a non-general-relativistic framework. The
safest way of deriving the deviations from a general relativistic
description of the geocentric motion of a satellite is to start from the
$\overline\beta -\overline\gamma -\overline\delta$-modified {\it
barycentric} equations of motion of the Earth and the satellite and to
take their difference. We restrict our attention to the effects
associated with the satellite (label 1), the Earth (label 2) and the Sun
(label 3), and introduce the notation\footnote{We do not need a special
notation for the mass of the satellite as it disappears  from the final
equations.}
${\bf r} \equiv r\, {\bf n} \equiv {\bf x}_{12} \equiv {\bf x}_1 -{\bf
x}_2$, ${\bf v}\equiv \dot{\bf r}$,
$D{\bf N} \equiv {\bf x}_{32} \equiv {\bf x}_3- {\bf x}_2$, ${\bf V}
\equiv \dot {\bf x}_{32} \equiv {\bf v}_3 - {\bf v}_2$, $m\equiv m_2$
and $M\equiv m_3$.

The general relativistic equations of motion exhibit some remarkable
features called ``effacement properties'' in Ref.~\cite{D87}. There is
an effacement of the internal structures of the gravitating bodies in
the sense that, to a high accuracy, one can write the equations of
motion in terms of only some ``centers of mass'' and some ``masses''.
There is also an effacement of the external universe in the dynamics of
a local gravitating system. For instance, after introducing some special
geocentric coordinate frame, the influence of the Sun on the dynamics
of an Earth satellite is effaced up to terms which decrease like
$D^{-3}$ when $D\to \infty$. Both types of effacement properties are
violated in tensor-scalar theories. The violation of the first type of
effacement properties shows up clearly in the Lagrangian (\ref{eq:1.1})
through the composition dependence of the effective coupling constant
$G_{AB}$. The violation of the second type of effacement properties
shows up in the fact that the geocentric equations of motion of a
satellite will contain terms proportional to $GM/c^2D$ and $GM/c^2D^2$
(with coefficients linear in $\overline\beta$, $\overline\gamma$ and
$\overline\delta$) which cannot be eliminated by a coordinate transformation.
In this paper, we shall be especially interested in the leading $(\propto
GM/c^2D)$ and subleading $(\propto GM/c^2D^2)$ violations of the effacement
of external influences appearing in non-Einsteinian theories. We shall
neglect the terms containing a factor $GM/c^2D^3$ which are, roughly
speaking, non-Einsteinian relativistic corrections to the Newtonian
tidal forces and are therefore extremely small\footnote{Compared to the
leading Newtonian acceleration, they are typically smaller by a factor
$\lesssim \overline\gamma\, (V/c)^2\, Mr^3/mD^3 \lesssim 5\times 10^{-19}
(a/R_{\rm Earth})^3$.}.

It is convenient to simplify the geocentric equations of motion by
rescaling the local spatial coordinates according to\footnote{This
rescaling is the $\overline\gamma$-dependent part of a standard
rescaling in the PPN literature \cite{W92,S89}. Its effect is to efface
the term proportional to $\overline\gamma GM /c^2D$ in the spatial
metric $g_{ij}$.}
\begin{equation}
 {\bf r}^{\rm new} = \left( 1 + \overline\gamma\ {GM\over c^2D}\right)
 {\bf r}^{\rm old}\ . \label{eq:2.1}
\end{equation}
 After  this rescaling, the geometric equations of motion have the form
\begin{equation}
 \ddot{\bf r} = {\bf A}_{\rm GR} + {\bf B} \label{eq:2.2}
\end{equation}
with
\begin{eqnarray}
 {\bf A}_{\rm GR} = - {Gm\over r^2}\, {\bf n} &-& {GM\over D^3}\,
 [ {\bf r} -3 ({\bf r} \cdot {\bf N}) {\bf N} ] \nonumber \\
 &+& O \left({GM\over D^4}\right) + O\left( {1\over c^2}\right)\ ,
  \label{eq:2.3}
\end{eqnarray}
denoting the general relativistic acceleration term (see \cite{DSX} for
the full expression of ${\bf A}_{\rm GR}$ including a relativistic
treatment of tidal forces), and
\begin{equation}
 {\bf B} = {\bf B}_0 + {\bf B}_1 + {\bf B}_2 + O \left[ {GM\over D^3}
  \left( {\overline\beta\over c^2} + {\overline\gamma\over c^2} +
   \overline\delta \right) \right] \label{eq:2.4}
\end{equation}
denoting the expansion of the non-Einsteinian acceleration terms in
successive powers of $D^{-1}$:
\begin{eqnarray}
 {\bf B}_0 &=& {Gm\over r^2}\, {\bf n} \left[ 2(\overline\beta +
  \overline\gamma) {Gm\over c^2r} - \overline\gamma\, {v^2\over c^2}
  - \overline\delta_1 - \overline\delta_2 \right] \nonumber \\
  &&+ {2\overline\gamma\over c^2}\, {Gm\over r^2}\,
     ({\bf n} \cdot {\bf v}) {\bf v}\ , \label{eq:2.5} \\
 {\bf B}_1 &=& (4\overline\beta - \overline\gamma)
    {Gm\over c^2D}{Gm\over r^2}\, {\bf n}\ , \label{eq:2.6} \\
 {\bf B}_2 &=& {GM\over D^2}\ {\bf N} \left[ -2\overline\beta
  {Gm\over c^2r} +\overline\delta_1 -\overline\delta_2 \right]\nonumber\\
  &&+ 2(\overline\beta +\overline\gamma) {GM\over D^2} {Gm\over c^2r}
   ({\bf n}\cdot {\bf N}) {\bf n} \nonumber \\
  &&+ {\overline\gamma\over c^2} {GM\over D^2} \left[ {\bf v}^2 {\bf N}
   - 2 ({\bf v} \cdot {\bf N}) {\bf v} \right] \nonumber \\
  &&+ 2 {\overline\gamma\over c^2} {GM\over D^2} ({\bf N}\times {\bf V})
  \times {\bf v} \ . \label{eq:2.7}
\end{eqnarray}

 To illustrate the non-Einsteinian terms proportional to $GM/D^3$ we can
quote its numerically dominant contribution:
\begin{equation}
 {\bf B}_3 = - (\overline\delta_1 +\overline\delta_3) {GM\over D^3}
   [{\bf r} -3 ({\bf r}\cdot {\bf N}) {\bf N}] + \cdots \label{eq:2.8}
\end{equation}
Let us also note that we shall neglect in this work the non-Einsteinian
acceleration terms depending on the higher (mass and spin) multipole
moments of the Earth. [We assume however that the Einsteinian effects
of these is fully taken into account in ${\bf A}_{\rm GR}$.] For instance,
the spin-orbit interaction adds a term of the form ${\bf B}_S = {\bf v}
\times {\bf H}_{\overline\gamma}$ to ${\bf B}$.
Here ${\bf H}_{\overline\gamma} = - {\overline\gamma\over c^2}
\mbox{\boldmath$\nabla$} \times (G{\bf S}_2 \times {\bf r}/r^3)$
where ${\bf S}_2$ is the spin angular momentum of the Earth, so that
\begin{equation}
{\bf B}_S = \overline\gamma {G\over c^2r^3}
\Biglb[{\bf v} \times {\bf S}_2
-3({\bf n}\cdot{\bf S}_2){\bf v}\times{\bf n}
\Bigrb]
\ . \label{eq:2.X}
\end{equation}
The spin-orbit acceleration ${\bf B}_S$ is smaller than the
velocity-dependent terms in ${\bf B}_0$, Eq.~(\ref{eq:2.5}), by a factor
$S_2/(mrv) \sim 0.02\, (a/R_{\rm Earth})^{-1/2}$. The other multipole
contributions will contain even smaller factors. They can be therefore
safely neglected with respect to the leading $\overline\gamma$- and
$\overline\beta$-dependent terms kept in
Eqs.~(\ref{eq:2.5})-(\ref{eq:2.7}).

 The last term on the right-hand side of Eq.~(\ref{eq:2.7}) has the form
of a Coriolis force, $2 \mbox{\boldmath$\Omega$}_{\overline\gamma} \times
{\bf v}$, with
\begin{equation}
  \mbox{\boldmath$\Omega$}_{\overline\gamma} = \overline\gamma
  {GM\over c^2D^2}\, {\bf N} \times {\bf V} \ . \label{eq:2.9}
\end{equation}
 As first discovered by De Sitter a similar relativistic Coriolis term,
$2 \mbox{\boldmath$\Omega$}_{\rm GR} \times {\bf v}$, with
\begin{equation}
  \mbox{\boldmath$\Omega$}_{\rm GR} \simeq {3\over 2} {GM\over c^2D^2}\,
  {\bf N} \times {\bf V} \label{eq:2.10}
\end{equation}
is present in Einstein's theory if one works in a geocentric frame which
has a fixed orientation with respect to the barycentric frame.
Alternatively, one can efface away the total relativistic Coriolis force
{}from the local equations of motion by working in a geocentric frame which
rotates with angular velocity
\begin{equation}
 \mbox{\boldmath$\Omega$}_{\rm tot} \equiv
\mbox{\boldmath$\Omega$}_{\rm GR}
 + \mbox{\boldmath$\Omega$}_{\overline\gamma} \simeq \left( 1 + {2\over 3}
  \overline\gamma \right) \mbox{\boldmath$\Omega$}_{\rm GR}
\label{eq:2.11}
\end{equation}
with respect to the barycentric frame. [See \cite{DSX} for a more
accurate treatment of $\mbox{\boldmath$\Omega$}_{\rm GR}$ than the one given
here.] Let us note that the rotation $\mbox{\boldmath$\Omega$}_{\rm tot}$
(``precession of the local Earth inertial frame'') is acting upon all
the satellite orbits (including the Moon), as well as upon the solid
Earth.  Therefore $\mbox{\boldmath$\Omega$}_{\rm tot}$ essentially
disappears in the position measurements effected by means of GPS or
Laser techniques. It appears  however when connecting local measurements
to global ones, e.g. through the use of Very Large Baseline Interferometry
(VLBI) techniques. It appears also in the Lunar Laser Ranging (LLR) data
because the internal dynamics of the Earth-Moon system is strongly
influenced by the Sun. The measurement of $\mbox{\boldmath$\Omega$}_{\rm
tot}$ through LLR data is now reaching the 1\% level (more precisely
$(2/3) \overline\gamma = -0.3 \pm 0.9\%$ \cite{LLR}). As the effects of
$\mbox{\boldmath$\Omega$}_{\rm tot}$ on low satellite orbits are
more difficult to observe\footnote{They correspond to spatial
displacements $\Omega_{\rm tot} a = (1+{2\over
3}\overline\gamma)(a/R_{\rm Earth})\times 59\ {\rm cm\ yr}^{-1}$.} this
does not look as a promising way of getting improved tests of
$\overline\gamma$. Therefore, in the following, we shall concentrate on
the other non-Einsteinian effects
${\bf B}$ in Eq.~(\ref{eq:2.2}). In other words, we shall study the
satellite motion in a geocentric frame precessing with the velocity
(\ref{eq:2.11}), so that the last term in Eq.~(\ref{eq:2.7}) is absent.

In such a dynamically non rotating geocentric frame one can verify that
the equations of motion of a satellite derive from a Lagrangian of the
form
\begin{equation}
  L = L_{\rm GR} + R \ , \label{eq:2.12}
\end{equation}
\begin{eqnarray}
 &&L_{\rm GR} = {1\over 2} {\bf v}^2 + {Gm\over r} - {GM\over 2D^3}
 [{\bf r}^2 - 3 ({\bf N}\cdot {\bf r})^2] \nonumber\\
 &&+\text{ other Newtonian and post-Newtonian contributions},
\nonumber\\
\label{eq:2.13}
\end{eqnarray}
\begin{equation}
 R = R_0 +R_1 +R_2 +O \left[ {GM\over D^3} \left({\overline\beta\over c^2}
+ {\overline\gamma\over c^2}+\overline\delta \right)\right]\ ,
\label{eq:2.14}
\end{equation}
\begin{eqnarray}
 R_0 &=& \left[ \overline\delta_1 + \overline\delta_2 + \overline\gamma
 {{\bf v}^2\over c^2} - \overline\beta \, {Gm\over c^2r} \right]
 {Gm\over r} \ , \label{eq:2.15} \\
 R_1 &=& - (4\overline\beta - \overline\gamma) {GM\over c^2D}\,
  {Gm\over r}\ , \label{eq:2.16} \\
 R_2 &=& \left( \overline\gamma {{\bf v}^2\over c^2} - 2 \overline\beta \,
  {Gm\over c^2r} + \overline\delta_1 - \overline\delta_2 \right)
 {GM\over D^2} ({\bf N} \cdot {\bf r})\ . \label{eq:2.17}
\end{eqnarray}
%%%%%%%%%%%%%%%%%%%%%%%%%%%%%%%%%%%%%%%%%%%%%%%%%%%%%%%%%%%%%%%%%%%%%%
\section{Non-Einsteinian effects in the motion of satellites}
\label{sec:3}

As we are discussing small deviations from a general relativistic
motion, we can discuss independently the perturbations depending upon
different powers of $D^{-1}$ in Eq.~(\ref{eq:2.14}) and superpose
linearly their effects.

The terms which are independent from the presence of the Sun,
Eqs.~(\ref{eq:2.5}) and (\ref{eq:2.15}), are  well known. Apart from a
constant renormalization of $G\to G_{\rm effective} \equiv G
[1+\overline \delta_1 +\overline\delta_2]$ (whose possible dependence
upon the satellite's composition is too small to be of observational
significance), they correspond to a geodesic motion in an
Eddington-modified Schwarzschild metric. The corresponding solution can
be written down explicitly (see e.g.
\cite{DD85} and Appendix A.2.3 of \cite{S89} for a simple form of the
explicit solution at the post-Newtonian accuracy). Short-period
perturbations correspond to spatial displacements of order
$(\overline\beta +\overline\gamma) Gm/c^2$, i.e. smaller than about
$10^{-3}$cm given the existing limits on $\overline\beta$ or
$\overline\gamma$. This is too small to be of observational significance.
Finally, the only observationally significant effect coming from ${\bf
B}_0$ is the {\it secular} advance of the perigee
\begin{equation}
  \delta_0 \omega = (2\overline\gamma -\overline\beta)\,
          {Gm\over c^2a(1-e^2)}\, nt \label{eq:3.1}
\end{equation}
where $a$ denotes the semi-major axis of the orbit, $e$ its eccentricity
and $n\equiv 2\pi /P \simeq (Gm/a^3)^{1/2}$ the orbital frequency [The
index zero on the left-hand side of (3.1) refers to the cause ${\bf
B}_0$ of this effect]. In terms of the ratio $\widehat a \equiv a/R$
where $R=6.371 \times 10^8$cm is the Earth radius, one
finds\footnote{Numerically $P= \widehat a^{3/2} \times 1.406$~h,
$Gm/c^2 = 0.4435$~cm, $Gm/Rc^2 =6.96 \times 10^{-10}$.}
\begin{equation}
  a\delta_0 \omega \simeq (2\overline\gamma -\overline\beta)
 {\widehat a^{-3/2}\over 1-e^2} \times \left( {t\over 1\ \rm{yr}}\right)
\times
  1.74 \times 10^4 \text{cm}\ . \label{eq:3.2}
\end{equation}

The terms which are proportional to $GM/D$, Eqs.~(\ref{eq:2.6}) and
(\ref{eq:2.16}) are, in principle, well known. They can be simply
interpreted as associated with a renormalization of the local
(geocentric) value of $G$ due to the proximity of the Sun: $G \to
G_{\text{loc}} = (1 -\eta GM/c^2 D) G$ \cite{W92}. The existence of this
(time-dependent) renormalization was taken into account in the
treatment of non-Einsteinian effects in Lunar Laser Ranging \cite{N73}
but was found to lead to a very small perturbation of the (directly
observed) radial distance (about 140 times smaller than the effects
associated with the violation of the equivalence principle,
$\overline\delta_1 - \overline\delta_2 \not= 0$ in Eq.~(\ref{eq:2.7})).
This conclusion is correct in the case of the Moon, but the main point
of the present paper is to emphasize that the situation is very different
in the case of low Earth satellites for two different reasons: (a) their
proximity to the Earth enhances the perturbation, and (b) the fact that
their motion can be tracked in three dimensions gives access to the
longitudinal displacement which happens to be enhanced by a large factor
$\propto n/n_2$, where $n_2\equiv (GM/a_2^3)^{1/2}$ denotes the orbital
frequency of the Earth around the Sun.

 It is possible to solve for the effects of $R_1$, Eq.~(\ref{eq:2.16}),
by the following method. Let us introduce new space and time
variables\footnote{We do not consider ${\bf r}'$ and $t'$ as new
coordinates in space-time but only as auxiliary variables which are
helpful to solve the equations of motion. Note that $t'$ would have
bad properties as a time coordinate as it would introduce a term linear
in $\eta GM/c^2D$ in the metric component $g'_{00}$.}
\begin{mathletters}
\label{eq:3.3}
\begin{eqnarray}
 {\bf r}' &=& {\bf r} - \mbox{\boldmath$\varepsilon$} = \left(
  1 - \eta \, {GM\over c^2D} \right)\, {\bf r}\ , \label{eq:3.3a} \\
  t' &=& t - \varepsilon^0 =
  t - 2\eta \int dt\, {GM\over c^2D}\ . \label{eq:3.3b}
\end{eqnarray}
\end{mathletters}
The difference between the Lagrangian $L'$ describing the motion in the
new variables and the original Lagrangian $L$ is given by \cite{DS85}
\begin{equation}
 L' - L = {\delta L\over \delta {\bf r}}
 (\mbox{\boldmath$\varepsilon$} - \varepsilon^0 {\bf v}) +
 {d\over dt}\, Q (\varepsilon)\ , \label{eq:3.4}
\end{equation}
where $\delta L/\delta {\bf r} = - (\ddot{\bf r} + Gm\,{\bf n} /r^2)
+ O(GM/D^3) +O(1/c^2)$ and where $dQ/dt$ is some total time derivative.
Using the identities
\begin{eqnarray}
 {\bf r} \cdot \left( \ddot{\bf r} + Gm\, {{\bf n}\over r^2} \right)
  &\equiv & - {\bf v}^2 + {Gm\over r} + {1\over 2}\,{d^2{\bf r}^2\over
  dt^2} \ , \nonumber \\
 {\bf v} \cdot \left( \ddot{\bf r} + Gm\, {{\bf n}\over r^2} \right)
  &\equiv & {d\over dt} \left( {1\over 2} {\bf v}^2 - {Gm\over r}\right)
   \ ,     \label{eq:3.5}
\end{eqnarray}
and operating by parts one finds
\begin{eqnarray}
 L' - L = \eta {GM\over c^2D}\,{Gm\over r} &-& {1\over 2}\eta {\bf r}^2
   {d^2\over dt^2} \left( {GM\over c^2D} \right) + {d\over dt}\, Q'
 (\varepsilon) \nonumber \\
 &+& O \left( {GM\over D^3} \varepsilon \right)
  + O\left( {\varepsilon\over c^2} \right) \ . \label{eq:3.6}
\end{eqnarray}
The first term on the right-hand side of (\ref{eq:3.6}) cancels the
contribution $R_1$ to $L$, while the second term is easily seen to be of
order $\eta (V/c)^2 GMr^2/D^3$, i.e. of order of the non-Einsteinian
relativistic corrections to tidal effects (see ${\bf B}_3$,
Eq.~(\ref{eq:2.8})) that we are neglecting in this work.

 The conclusion is that the effect of $R_1$ is equivalent to performing
the transformation (\ref{eq:3.3}) on the solution ${\bf r}' (t')$ of the
equations of motion in which $R_1$ (and ${\bf B}_1$) have been set equal
to zero. Note that this method can be applied either to the full term
$GM/c^2D$ or only to its time-varying piece due to the eccentricity
$e_2$ of the Earth orbit around the Sun (we neglect $(e_2)^2$ for
simplicity):
\begin{equation}
 \widetilde{GM\over c^2D} \equiv {GM\over c^2D} - {GM\over c^2a_2} \simeq
 e_2 {GM\over c^2a_2}\, \cos\, n_2 (t-t_2) \ ,  \label{eq:3.7}
\end{equation}
where $t_2$ is the time of passage of the Earth at its perihelion. The
time-independent renormalization of $G$ associated with $GM/c^2a_2$ is
locally unobservable. The time-dependent perturbations in the spatial
positions, Eq.~(\ref{eq:3.3a}), are of order $0.1 \eta\,\hat a$~cm,
i.e. too small to be observable. Finally, the only observationally
significant effect coming from ${\bf B}_1$ is the change, given by
Eq.~(\ref{eq:3.3b}), in the time according to which the satellite runs
on its orbit. If we introduce the mean anomaly $\ell$, i.e. an angle
connected with the position on the orbit which, in the unperturbed
motion, varies linearly in time, we shall have in the perturbed motion
\begin{equation}
 \ell = n\, t' + {\rm const}= n \left( t-2\eta \int dt\,
\widetilde{GM\over
 c^2D} \right) + {\rm const} \ , \label{eq:3.8}
\end{equation}
or $\ell = nt + \delta_1 \ell$ with
\begin{equation}
 \delta_1 \ell = - 2\eta\, e_2\, {GM\over c^2a_2}\, {n\over n_2}\,
  \sin\, n_2 (t-t_2) + {\rm const}\ .   \label{eq:3.9}
\end{equation}
Using $e_2 = 1.673 \times 10^{-2}$ and $GM/c^2a_2 = 9.87 \times
10^{-9}$, the perturbation (\ref{eq:3.9}) corresponds to an along-track
displacement
\begin{equation}
 a\delta_1 \ell = - (\overline\beta - {1\over 4} \overline\gamma)
 \hat a^{-1/2} \sin \left(2\pi\, {t-t_2\over 1\ {\rm yr}}\right) \times
 5.25 \times 10^3\ \text{cm} \ . \label{eq:3.10}
\end{equation}

The yearly modulation of $G_{\rm loc} = (1 -\eta\, GM/c^2D) G$ has
other consequences which, in principle, enter satellite data through
the motion of, e.g., Laser stations on the ground used to track a
satellite. First, the radius of the Earth will undergo a corresponding
yearly modulation. Using $\partial \ln R/\partial \ln G \simeq
-0.1 $\cite{MD64}, the amplitude of this modulation is
\begin{eqnarray}
 \delta_1 R &\simeq & -0.1\,R\, {\delta G_{\rm loc}\over G_{\rm loc}}
\simeq
 0.1\,R\,e_2\, {4GM\over c^2a_2} \left( \overline\beta - {1\over 4}
 \overline\gamma \right) \cos n_2 (t-t_2)\nonumber\\
 &\simeq & \left( \overline\beta - {1\over 4} \overline\gamma \right)
 \cos \left( 2\pi\, {t-t_2\over 1\ {\rm yr}}\right) \times 4.21 \times
10^{-2}
 \, {\rm cm}\ ,
\label{eq:3.11}
\end{eqnarray}
which is negligibly small. Second, this yearly breathing of the Earth
will entail a corresponding modulation of its angular velocity, say
$\Omega_2$. Using $\delta\Omega_2/\Omega_2 = - \delta {\cal I}_2/{\cal
I}_2$ and $\delta {\cal I}_2/{\cal I}_2 \simeq -0.17\, \delta G/G$
\cite{LF78} where ${\cal I}_2$ denotes the Earth inertia moment, this
leads, by integration, to a yearly modulation of the angle of rotation of
the Earth and therefore a corresponding longitudinal displacement of
ground stations:
\begin{eqnarray}
 R\delta_1 \varphi &\simeq& -0.17\,R\, {\Omega_2\over n_2}\, e_2\,
  {4GM\over c^2a_2} \left( \overline\beta - {1\over 4}
 \overline\gamma \right) \sin n_2 (t-t_2)\nonumber\\
 &\simeq & -\left( \overline\beta - {1\over 4} \overline\gamma \right)
\sin \left( 2\pi\, {t-t_2\over 1\ {\rm yr}}\right) \times 26.1\ {\rm cm}
\ . \label{eq:3.12}
\end{eqnarray}
Again, this effect is negligible compared to the orbital effect
(\ref{eq:3.10}). [It would moreover probably be impossible to distinguish
{}from other yearly modulations in the rotation of the Earth.]

 It remains to study the orbital effects of the terms proportional
to $GM/c^2D^2$ in the equations of motion. One checks that the
short-period effects associated with ${\bf B}_2$, Eq.~(\ref{eq:2.7}), are
too small to be observationally relevant. One can get the secular
effects of the orbital elements caused by ${\bf B}_2$ by averaging over
one orbital period the time derivatives of the energy, the angular
momentum and the Lagrange-Laplace (-Runge-Lenz) vector. One finds that
$\langle da/dt
\rangle_2 =0$ and
\begin{mathletters}
\label{eq:3.13}
\begin{eqnarray}
 \langle d{\bf e}/dt\rangle_2 &=& {\bf f}_2\times
\mbox{\boldmath$\ell$}\ ,
         \label{eq:3.13a}\\
 \langle d\mbox{\boldmath$\ell$}/dt\rangle_2 &=& {\bf f}_2\times
{\bf e}\ ,
          \label{eq:3.13b}
\end{eqnarray}
\end{mathletters}
where ${\bf e}$ denotes the eccentricity vector of the orbit (i.e. a
vector of magnitude $e$ directed toward the perigee),
$\mbox{\boldmath$\ell$}\equiv
\sqrt{1-e^2}\, {\bf c}$, with ${\bf c}$ a unit vector along the orbital
angular momentum, and where the average ``forcing term'' caused by ${\bf
B}_2$ is
\begin{mathletters}
\label{eq:3.14}
\begin{eqnarray}
 {\bf f}_2 &=& k\, {GM\over D^2}\, {\bf N} \ , \label{eq:3.14a}\\
 k &\equiv & {3\over 2na} \left( \overline\delta_1 - \overline\delta_2
      \right) - {na\over 2c^2}\, (4\overline\beta -\overline\gamma)\ .
 \label{eq:3.14b}
\end{eqnarray}
\end{mathletters}
In the general case of eccentric orbits, Eq.~(\ref{eq:3.13b}) shows that
${\bf B}_2$ will produce variations of the inclination $I$ and the node
$\Omega$ to the orbit of the satellite. Here, we shall concentrate on
the more usual case where the eccentricity is small. Then,
Eq.~(\ref{eq:3.13a}) with $\mbox{\boldmath$\ell$} \simeq {\bf c}$
gives the contribution
of ${\bf B}_2$ to the variation of ${\bf e}$. We must add the Newtonian
effects of the quadrupole moment of the Earth ($J_2 = 1.08263\times
10^{-3}$). Let us introduce the basis $({\bf a},{\bf b})$ of the orbital
plane where
${\bf a}$ is directed toward the ascending node and ${\bf a} \times {\bf
b} = {\bf c}$. Because of the Newtonian effects of $J_2$ this basis is
rotating around the Earth's polar axis with angular velocity $\dot\Omega
\simeq - {3\over 2} n\, J_2 \hat a^{-2} \cos I$ (see e.g. \cite{DEF94}).
Let $d'/dt$ denote a time derivative in the rotating frame $({\bf a},
{\bf b}, {\bf c})$. We get the following equation for the secular time
evolution of the two independent components of ${\bf e}$ with respect to
the vectors $({\bf a}, {\bf b})$
\begin{equation}
 d'{\bf e} /dt = {\bf c} \times [\dot\omega_N\, {\bf e} -
{\bf f}_\bot (t)]
  \ , \label{eq:3.15}
\end{equation}
where ${\bf f}_\bot \equiv ({\bf f}_2 \cdot {\bf a}) {\bf a} + ({\bf f}_2
 \cdot {\bf b}) {\bf b}$ denotes the projection of ${\bf f}_2$ onto
the orbital plane, and where
\begin{eqnarray}
\dot\omega_N &=& {3\over 4}\,{nJ_2\over \hat a^2}\, (4-5\,\sin^2I)
\nonumber\\
 &\simeq & (4-5\,\sin^2I) \hat a^{-7/2}\times 2\pi/(0.20\ {\rm yr})
\label{eq:3.16}
\end{eqnarray}
is the Newtonian perigee advance due to the Earth's quadrupole moment
\cite{DEF94}. Eq.~(\ref{eq:3.15}) constitutes an inhomogeneous linear
differential equation in ${\bf e} =e \cos \omega\, {\bf a} + e\sin
\omega\, {\bf b}$. The time dependence of the forcing term ${\bf f}_\bot
(t)$ comes from a combination of the yearly variation of ${\bf f}_2$,
Eq.~(\ref{eq:3.14a}), with the rotation of the orbital plane with angular
velocity $\dot\Omega$ around the Earth's polar axis. Following our recent
work \cite{DEF94}, we can easily solve for ${\bf e}(t)$ by decomposing
${\bf f}_\bot$ in a sum of constant-norm vectors ${\bf f}_\alpha$
rotating with constant angular frequencies $\dot\omega_\alpha$ in the
$({\bf a},{\bf b})$-plane: ${\bf f}_\bot = \Sigma_\alpha {\bf f}_\alpha$.
Alternatively, one can work with complex numbers, $x\, {\bf a} +y\, {\bf
b} \to z \equiv x+iy$, and decompose ${\bf f}_\bot \cdot ({\bf a} +i{\bf
b})$ into harmonic components $\propto \exp (i\dot\omega_\alpha t)$. Then
the general solution of (\ref{eq:3.15}) has the form
\begin{equation}
 {\bf e}(t) ={\bf e}_N (t)+\sum_\alpha {\bf e}_\alpha (t)\ ,\label{eq:3.17}
\end{equation}
where
\begin{equation}
 {\bf e}_\alpha \equiv {{\bf f}_\alpha\over \dot\omega_N -\dot\omega_\alpha}
  \ , \label{eq:3.18}
\end{equation}
and where ${\bf e}_N(t)$ is a constant-norm vector rotating in the
$({\bf a},{\bf b})$-plane with angular velocity $\dot\omega_N$
(homogeneous solution of (3.15)). In terms of complex numbers, $z={\bf
e}\cdot ({\bf a} + i{\bf b}) = e\, \exp i\omega$, this general solution
reads
\begin{eqnarray}
 [e(t) \exp i\omega (t)]_2 =&& e_N \exp i(\dot\omega_Nt
+\sigma_N)\nonumber\\
 &&+ k{GM\over 4D^2}
\left[ {(1+\cos I_2)(1+\cos I)\over \dot\omega_N +
    \dot\Omega -n_2} \exp +i (n_2t - \Omega) \right. \nonumber \\
 && \qquad + {
 (1+\cos I_2)(1-\cos I)\over \dot\omega_N -
    \dot\Omega +n_2} \exp -i (n_2t - \Omega) \nonumber \\
 && \qquad + {
 (1-\cos I_2)(1+\cos I)\over \dot\omega_N +
    \dot\Omega +n_2} \exp -i (n_2t + \Omega) \nonumber \\
 && \qquad + {
 (1-\cos I_2)(1-\cos I)\over \dot\omega_N -
    \dot\Omega -n_2} \exp +i (n_2t + \Omega) \nonumber \\
 && \qquad + {2\sin I_2 \sin I\over \dot\omega_N -n_2} \exp +i\, n_2 t
     \nonumber \\
 && \left. \qquad + {2\sin I_2 \sin I\over \dot\omega_N +n_2} \exp - i\,
      n_2 t \right]\ , \label{eq:3.19}
\end{eqnarray}
where $I_2 = 23.45^{\circ}$ is the inclination of the apparent orbit of
the Sun with respect to the Earth equatorial plane, and where we
have neglected the eccentricity $e_2$ of the Earth's orbit.

We see clearly from Eq.~(\ref{eq:3.19}) that the effects of ${\bf B}_2$
can be enhanced by small divisors if the semi-major axis and the
inclination of the satellite orbit (on which both $\dot\omega_N$ and
$\dot\Omega$ depend) are appropriately chosen. The locus of these
resonances in the $(a,I)$ plane is the same as for the perturbations
due to the orbital velocity of the Earth around the Sun in
non-boost-invariant gravity theories. It will be found in Fig.~3 of
\cite{DEF94}. Note that these resonances exist only if $a$ is smaller
than $2.65 R$. Because of the smallness of
$I_2$, the dominant contributions in the square brackets on the
right-hand side of Eq.~(\ref{eq:3.19}) are the first two ones
corresponding to the frequencies $\pm (n_2 -\dot\Omega)$ (see the solid
lines in Fig.~3 of \cite{DEF94}). However, the price to pay to take
advantage of these resonances is to dispose of a long time of
observation. In this work we have in mind using tracking data from
missions dedicated to other purposes (e.g. GPB or STEP) which may last
only for a year or so. Then, the best situation is when $\dot\omega_N \pm
(\dot\Omega -n_2)$ is of order $2\pi /1$~yr. In such a case, we see from
Eq.~(\ref{eq:3.19}) that ${\bf B}_2$ will cause spatial displacements
with periods $\sim 1$~yr and with amplitudes
\begin{eqnarray}
a\delta_2 e &\sim& ak {GM\over 2D^2} {1\pm \cos I\over
    \dot\omega_N \pm (\dot\Omega -n_2)} \nonumber \\
 &\sim & [1.80\times 10^9\ \hat a^{3/2} (\overline\delta_1 -
\overline\delta_2) -
1.67\ \hat a^{1/2} (\overline\beta - {1\over 4} \overline\gamma)]
\nonumber\\
 &&\qquad \times (1\pm \cos I)\ {2\pi /1\ {\rm yr}\over \dot\omega_N
\pm (\dot\Omega -n_2)}
 \ \text{cm} \ , \label{eq:3.20}
\end{eqnarray}
where $\overline\delta_1 - \overline\delta_2 \simeq
\widehat\delta_1 -\widehat\delta_2 + 1.86\times
10^{-9} (\overline\beta - \overline\gamma/4)$, cf. Eq.~(\ref{eq:1.3}).
For low Earth satellites $(\hat a\sim 1)$ this is much smaller than the
effects of ${\bf B}_0$ and ${\bf B}_1$ discussed above, and will
therefore not contribute much to the measurability of $\overline\beta$,
$\overline\gamma$ and $\overline\delta$. By comparing however the
negative power law dependences upon $\hat a$ of Eqs.~(\ref{eq:3.2}) and
(\ref{eq:3.10}) with the corresponding positive power dependences in
Eq.~(\ref{eq:3.20}) we see in retrospect why the most sensitive
non-Einsteinian test in the motion of the Moon $(\hat a\simeq 60)$ can
be the violation of the equivalence principle (term $\propto
\overline\delta_1 - \overline\delta_2$ in $k$ \cite{N68,N73}).

 Finally, let us consider for completeness the orbital effects of a slow
time variation of $G$ and/or the masses. Going back to the level of the
original Lagrangian (\ref{eq:1.1}) we must consider that the mass of
the satellite $m_s$, the mass of the Earth $m$ and $G$ can vary in the
Newtonian-approximation Lagrangian for the geocentric motion:
\begin{equation}
 L_N = {1\over 2} m_s (t) \left( {d{\bf r}\over dt}\right)^2 +
 {G(t) m_s(t) m(t)\over r} \ . \label{eq:3.21}
\end{equation}

We can solve for the dynamics of $L_N$ by a simple generalization of the
method used above for the yearly variation of $G_{\text{loc}}$.

Indeed, let us consider more generally a perturbed Lagrangian of the type
$L=L_0 +L_1$ with $L_0 ={1\over 2} {\bf v}^2 + Gm/r$ and $L_1 ={1\over 2}
a(t) {\bf v}^2 +b(t) Gm/r$ where $Gm$ is constant and where $a(t)$, $b(t)
\ll 1$. Let us introduce the following new space and time variables
\begin{mathletters}
\label{eq:3.22}
\begin{eqnarray}
 {\bf r}' &=& (1 + a+b) {\bf r} \ , \label{eq:3.22a}\\
      t'  &=& t + \int dt (a+2b)\ . \label{eq:3.22b}
\end{eqnarray}
\end{mathletters}
By using Eqs.~(\ref{eq:3.4}) and (\ref{eq:3.5}) we find that
\begin{equation}
 L' = L_0 + {1\over 2}\, {\bf r}^2 {d^2\over dt^2} (a+b) + {d\over dt}\,
  Q' \ . \label{eq:3.23}
\end{equation}
We can apply this result to the Lagrangian (\ref{eq:3.21}) decomposed
as $L_N = m_s [L_0 +L_1]$ by writing $m_s(t) =m_s (1+t\dot m_s/m_s)$,
$m(t) =m(1+t\dot m/m)$, $G(t) =G(1+t\dot G/G)$ (where $m_s$, $m$ and
$G$ denote some constants). [Here, one is making the assumption that the
characteristic time scale $T$ of variation of $G(t)$, $m_s(t)$, $m(t)$ is
very large so that one can work to first order in $T^{-1}$.] This yields
$a=t\dot m_s/m_s$ and $b= t(\dot G/G +\dot m/m + \dot m_s/m_s)$, and
therefore $(d^2/dt^2) (a+b) =0$ in Eq.~(\ref{eq:3.23}). Finally, we get
that the motion in terms of the variables
\begin{mathletters}
\label{eq:3.24}
\begin{eqnarray}
 {\bf r}' &=& \left[ 1 +t \left( {\dot G\over G} + {\dot m\over m} +
  2 {\dot m_s\over m_s} \right) \right] {\bf r} \ , \label{eq:3.24a}\\
   t' &=&  t + {1\over 2}\, t^2 \left( 2 {\dot G\over G} + 2{\dot m\over m}
   + 3 {\dot m_s\over m_s} \right)\ , \label{eq:3.24b}
\end{eqnarray}
\end{mathletters}
is described by the unperturbed Lagrangian $L_0$.

Knowing already that $|\dot G/G| < 10^{-11}$~yr$^{-1}$ and $|\dot m/m|
< 10^{-13}$~yr$^{-1}$ we see that the effects (\ref{eq:3.24a}) on the size
of the orbit is too small to be observable in satellite data. On the
other hand the effect (\ref{eq:3.24b}) leads to a mean anomaly
perturbation
\begin{equation}
 \ell = nt' + \text{const} = nt + {1\over 2} nt^2 \left( 2 {\dot G\over G}
  + 2 {\dot m\over m} + 3 {\dot m_s\over m_s} \right)\ , \label{eq:3.25}
\end{equation}
corresponding to an along-track displacement\footnote{The result
(\ref{eq:3.26}) assumes that one has experimentally access to the basic
dynamical time $t$. Taking into account that time is measured by atomic
clocks based on, e.g., a Bohr frequency $\propto m_e\alpha^2$ adds the
terms $-{1\over 2}\dot m_e/m_e - \dot\alpha /\alpha$ in the
parenthesis on the right-hand side of Eq.~(\ref{eq:3.26}).}
\begin{eqnarray}
 a\delta \ell &=& nat^2 \left( {\dot G\over G} + {\dot m\over m} +
 {3\over 2} {\dot m_s\over m_s} \right) \nonumber \\
 &\simeq & \hat a^{-1/2} \left( {\dot G\over G} + {\dot m\over m} +
  {3\over 2} {\dot m_s\over m_s} \right) t^2
 \times 2.50 \times 10^{13} \text{cm yr}^{-1}\ , \label{eq:3.26}
\end{eqnarray}
which suggests that the level $\dot G/G \sim 10^{-13}$~yr$^{-1}$ might be
reachable by satellite data [assuming that the ordinary sources of
variation of the mass of the Earth are smaller, or, at least, measurable
with better precision].
%%%%%%%%%%%%%%%%%%%%%%%%%%%%%%%%%%%%%%%%%%%%%%%%%%%%%%%%%%%%%%%%%%%%%%
\section{Conclusions}
\label{sec:4}

 We have reexamined the non-Einsteinian effects in the orbital motion
of low-orbit artificial Earth satellites. We worked within the most
conservative framework for alternative gravity theories, i.e. the
assumption that gravity contains, besides the usual Einsteinian tensor
interaction, a scalar contribution. The post-Newtonian limit of generic
tensor-scalar gravity theories exhibit three types of non-Einsteinian
effects: (i) deviations from the general relativistic post-Newtonian
effects, parametrized by two (Eddington) parameters $\overline\beta
\equiv \beta -1$, and $\overline\gamma \equiv \gamma -1$, (ii) combined
violations of the weak and strong equivalence principles, parametrized
by body-dependent parameters $\overline\delta_A \equiv
\widehat\delta_A + (4 \overline\beta - \overline\gamma)
E^{\text{grav}}_A / m_A c^2$; and (iii) possible slow time
variabilities of Newtonian parameters, measured by
$\dot G/G$ and $\dot m_A /m_A$.

 Let us recall that the Lunar Laser Ranging (LLR) experiment has, as
predicted \cite{N68,N73}, proven to be a superb tool for measuring
$\overline\delta_{\text{Moon}} - \overline\delta_{\text{Earth}} = (-2.7
\pm 6.2) \times 10^{-13}$, i.e. for testing the validity of the
effacement of the internal structure of the bodies in their
translational equations of motion. The present paper indicates that the
centimeter-level tracking of low-orbit drag-free satellites is a
promising tool for measuring at the $10^{-4}$ level two independent
combinations of $\overline\beta$ and $\overline\gamma$ $[\overline\beta
-2\overline\gamma$ Eq.~(\ref{eq:3.1}) and $\overline\beta - {1\over 4}
\overline\gamma$, Eq.~(\ref{eq:3.10})] and for probing the time
variability of $G$ (and the masses) at the $10^{-13}$~yr$^{-1}$ level,
Eq.~(\ref{eq:3.26}). It is interesting to notice that the satellite
measurement of $\eta \equiv 4 \overline \beta - \overline\gamma$ is
complementary to the LLR measurement of $\overline\delta_1 -
\overline\delta_2$ in two ways: (i) conceptually it is a test of the
effacement of the external universe in the equations of motion of a local
system, and (ii) technically it gives directly access to the parameter
$\eta$ which is ``contaminated'' by possible weak-equivalence-principle
violations in the LLR observable $\overline\delta_1 -\overline\delta_2
=\widehat\delta_1 -\widehat\delta_2 + \eta \times
4.45
\times 10^{-10}$
\cite{LLR}.

 A crucial difficulty in reaching the precision levels suggested by our
analysis will be to protect the motion of the satellite from
non-gravitational forces\footnote{One notes in particular that solar
radiation pressure effects mimic exactly the yearly variation of
$G_{\text{loc}}$ but are several orders of magnitude larger.}.
The most promising orbital effects, Eqs.~(\ref{eq:3.1}), (\ref{eq:3.10})
and (\ref{eq:3.26}), come from the build up of small effects over
observation times of the order of one year. The drag-free system of the
satellite must therefore be effective for correspondingly ultra low
frequencies. It may well be that the active drag compensation systems
of the type planned for the GPB satellite will fall short of providing
the needed protection. The best system would probably be a passive
drag-free system where the satellite floats freely within a large shell.

If it appears impossible to use the orbital data of drag-free satellites
conceived for other missions [like GPB or the Satellite Test of the
Equivalence Principle (STEP)], it may be worth thinking about a dedicated
mission [say the Satellite Test of the Effacement property
(STEFF)] consisting of an optimized passive drag-free satellite: a small
ball freely floating within a large shell which follows optically the
motion of the ball and which carries some thrusters and GPS or DORIS
receivers and/or laser corner cubes. The orbital data of such a mission
would be of value both for geodesy and as tests of relativistic gravity.

Other difficulties concern the knowledge of the multipole moments of
the Earth which limit other proposals (see \cite{C89}). For instance,
the present uncertainty in the Earth quadrupole $J_2$, $\delta J_2 /J_2
\sim 6 \times 10^{-7}$ \cite{MNF87} contributes an uncertainty in the
predicted (Newtonian) value of the perigee advance which corresponds
to the uninteresting level $\delta (\overline\beta - 2 \overline\gamma)
\sim 1$. This is however not a serious difficulty as the
high-precision satellite data that we wish to use to get (as by
products) measurements of $\overline\beta$ and $\overline\gamma$, will
be primarily used anyway to refine the measurement of the Earth gravity
field with a correspondingly high precision (see e.g. \cite{Tapley}).

One might think that the type of orbital tests we have been discussing
will soon loose their interest as other planned experiments aim at
reaching higher precisions in measuring possible deviations from General
Relativity. [We have in mind, in particular, Gravity Probe B (which
should measure $\overline\gamma$ at the $3\times 10^{-5}$
level \cite{BE94}) and the Solar Orbit Relativity Test (SORT) \cite{V93}
which aims at the $10^{-7}$ level for $\overline\gamma$.] As these tests
will not measure $\overline\beta$, we see on the contrary that orbital
tests may provide our best handle on the parameter\footnote{Let us
note in passing that the scenario of Ref.~\cite{DP} indicates that the
ratio $-\overline\beta /\overline\gamma = \beta_3/4$ might be of order
10 (if $\kappa \sim 1$).} $\overline\beta$.
Assuming the knowledge of $\overline\gamma$, the present analysis shows
that orbital data will contain two independent signals,
Eq.~(\ref{eq:3.1}) and (\ref{eq:3.10}), for measuring $\overline\beta$
at the $10^{-4}$ level.
One can note that while the perigee advance of Eq.~(\ref{eq:3.1}) is
difficult to measure in the usual case of low-eccentricity orbits, the
yearly modulation (\ref{eq:3.10}) is essentially independent of the
magnitude of $e$. It remains however to see to what extent it will
possible to distinguish it from the other yearly (gravitational)
perturbations in the orbit. [Note that the phase of the signal
(\ref{eq:3.10}) is predicted and that the period is the anomalistic year
(perihelion to perihelion).]
Let us also remark that, contrary to many of the effects discussed in
\cite{DEF94}, the magnitudes of the leading perturbations we have
discussed, Eqs.~(\ref{eq:3.1}), (\ref{eq:3.10}) and (\ref{eq:3.26}), are
independent of the value of the inclination of the orbit. A strong
dependence upon
$I$ appeared only in the subdominant effects caused by ${\bf B}_2$,
Eq.~(\ref{eq:3.19}).

Finally, let us stress that the value of the approximate analytical
solutions for non-Einsteinian perturbations that we have given above is
mainly indicative of the results one can expect. In practice, we advise
to resort to direct numerical integration of the equations of motion
(\ref{eq:2.2}). For ${\bf A}_{\text{GR}}$ one should use the full general
relativistic post-Newtonian equations of motion \cite{DSX}, to which one
can add linearly the effect of ${\bf B}$. The accuracy with which we
have given ${\bf B}$ in Eqs.~(\ref{eq:2.5})-(\ref{eq:2.7}) should be
amply sufficient. [If one wishes one can also consider the non-Einsteinian
effect associated with the spin angular momentum of the Earth ${\bf
S}_2$, i.e. add the term ${\bf B}_S$, Eq.~(\ref{eq:2.X}), to ${\bf B}$.]

%%%%%%%%%%%%%%%%%%%%%%%%%%%%%%%%%%%%%%%%%%%%%%%%%%%%%%%%%%%%%%%%%%%%%%

\end{document}